\documentclass[a4paper,12pt]{article}
\begin{document}
\vspace {1.5cm}
\begin{center}
\baselineskip24pt
\noindent {\bf Comment on "Chiral symmetry and the intrinsic structure of the
\newline
nucleon"}- by D.B.Leinweber,A.W Thomas and R.D.Young. \newline
\vskip 8 pt
G.Morpurgo
\vskip 5 pt
Universit\`a di Genova and INFN - Genova (Italy)
\end{center}
\vskip 5 pt
\baselineskip24pt
The authors of ref.[1], using QCD plus approximate chiral symmetry show that the
"perfect" prediction 3/2 for $\vert \mu(p)/\mu(n) \vert$ is coincidental. I 
agree; infact this was discussed in [2b,3,5] on the
basis of the general QCD parameterization (GP). The argument in [2,3] is
more general than that in [1], but [1] is of interest, because
it exemplifies an explicit mechanism that may produce the
coincidence. Yet I disagree with a statement in [1]. It is: "Within
the constituent quark model this ratio would remain constant at 3/2, independent
of the change of the quark mass". This is so for an additive [4], but not
for a general constituent model.

For the lowest hadronic multiplets the GP relates [2,3] all possible
constituent models to QCD. It parametrizes in the most general
way compatible with QCD the properties (masses, magnetic moments, etc.) of the
lowest hadron multiplets. It was started [2a] to explain the fair quantitative
success of the simple non relativistic model [4]. 
Although non covariant (we work in a given frame), the GP 
is relativistic, based only on general properties of relativistic QCD.
 Also [3] the renormalization point for quark masses can be selected at will
in the QCD Lagrangian; i.e. the GP is compatible [3] with a quasi-chiral
description, with light u,d quarks.(The script symbols for quark fields 
that I used in [2] may be confusing on this; they seem to imply $\approx 
"300"$ MeV masses for u,d quarks in the QCD Lagrangian. Standard
symbols u,d,s were used from [3] on.)

For each quantity considered, the GP gives the most general spin flavor 
parameterization compatible with QCD. This alone is not much. Indeed e.g. for 
the \textbf{8} plus \textbf{10} baryon masses, the GP has 8 parameters to fit
8 masses. Clearly trivial! But, fitting the data, a hierarchy in the parameters
emerges: The parameters multiplying spin-flavor structures of increasing
complexity are smaller and smaller. This is true for any quantity treated
so far, in particular the baryon magnetic moments. The reduction factor due to
increasing complexity of GP terms is, from the data, the product of $\cong 0.3$
for each flavor breaking factor (FB) and 0.2-0.37 for each pair of different
indices in the term ("gluon exchange" factor [2b]). In [2,3] we 
parametrized the magnetic moments of the octet baryons up to first order
in FB. There are seven data and seven parameters, called $g_{i}$
The $g_{i}$'s are functions of the quark masses and $\Lambda_{QCD}$.
 To first order FB the expression so obtained is the most general one in QCD
or in any constituent quark model compatible with it. It was underlined in
[2b,3,5] that all $g_{i}$'s agree, to a factor 2, with the expectations from 
the hierarchy taking a gluon exchange factor from 0.2 to 0.35; only
$g_{3}$ is $\approx7$ (or 10) times smaller ($g_{3}$ is,in the GP, the
coefficient of the term giving the deviation of $\vert\mu(p)/\mu(n) \vert$
from 3/2. In [(2b), Eq.(23)] such factor 7 (or 10) was possibly attributed to
chance.

 This conclusion on chance is much reinforced (Sect.4 of [5]) considering only
the parametrized $\mu$'s of the non strange baryons of \textbf{8}+\textbf{10}
 (p,n,$\Delta$'s). Then we have four parameters ($\alpha,\beta,\gamma,
\delta$). The "perfect" 3/2 arises from $g_{3}=\delta-\beta-4\gamma
\cong0$. That this particular combination almost vanishes can be due only to a
 chance cancellation; a cancellation compatible with the typical hierarchy
reduction, and with the known $\mu(p)$,$\mu(n)$ and $\mu(\Delta\to p\gamma)$.
A measurement of $\mu(\Delta^{+})$'s would allow a further check.

\end{document}